\documentclass[preprint2]{aastex}
\usepackage{graphics}

\newcommand{\microns}{\micron}
\advance \topmargin by -\headheight
\advance \topmargin by -\headsep
\evensidemargin \oddsidemargin
\begin{document}

\title{A Luminous Infrared Companion in the Young Triple System WL~20}

\author{Michael E. Ressler\altaffilmark{1,2,3}}

\altaffiltext{1}{Visiting Astronomer at the W.~M.~Keck Observatory, which is
operated as a scientific partnership among the California Institute of
Technology, the University of California and the National Aeronautics and
Space Administration. The Observatory was made possible by the generous
financial support of the W.~M.~Keck Foundation.} 

\altaffiltext{2}{Visiting Astronomer at the Infrared Telescope Facility, which
is operated by the University of Hawaii under contract from the National
Aeronautics and Space Administration.} 

\altaffiltext{3}{Observations with the Palomar 5-m telescope were obtained
under a collaborative agreement between Palomar Observatory and the Jet
Propulsion Laboratory.}

\affil{Jet Propulsion Laboratory\\
 4800 Oak Grove Drive, Pasadena, CA 91109\\ ~\\ ~}

\and{}

\author{Mary Barsony\altaffilmark{3}}

\affil{Space Science Institute\\
 3100 Marine Street, Suite A353, Boulder, CO 80303--1058}

\begin{abstract}
We present spatially resolved near-infrared and mid-infrared (1--25~\microns)
imaging of the WL~20 triple system in the nearby (\( d=125 \)~pc) \( \rho \)
Ophiuchi star-forming cloud core. We find WL~20 to be a new addition to the
rare class of {}``infrared companion systems{}'', with WL~20:E and WL~20:W
displaying Class II (T-Tauri star) spectral energy distributions (SEDs) and
total luminosities of 0.61 and 0.39~\( L_{\sun } \), respectively, and
WL~20:S, the infrared companion, with a Class I (embedded protostellar) SED
and a luminosity of 1.0--1.8~\( L_{\sun } \). WL~20:S is found to be highly
variable over timescales of years, to be extended (40~AU diameter) at
mid-infrared wavelengths, and to be the source of the centimeter emission in
the system.

The photospheric luminosities of 0.53~\( L_{\sun } \) for WL~20:E and 0.35~\(
L_{\sun } \) for WL~20:W, estimated from our data, combined with existing,
spatially resolved near-infrared spectroscopy, allow us to compare and test
current pre-main-sequence evolutionary tracks. The most plausible,
non-accreting tracks describing this system are those of \citet{dantona}.
These tracks give an age of 2--\( 2.5\times 10^{6} \)~yr and masses of
0.62--0.68~\( M_{\sun } \) for WL~20:E and 0.51--0.55~\( M_{\sun } \) for
WL~20:W, respectively. The age and mass of WL~20:S cannot be well determined
from the currently available data. WL~20:E and WL~20:W fall into the region of
the H-R diagram in which sources may appear up to twice as old as they
actually are using non-accreting tracks, a fact which may reconcile the
co-existence of two T-Tauri stars with an embedded protostar in a triple
system. The derived masses and observed projected separations of the
components of the WL~20 triple system indicate that it is in an unstable
dynamical configuration, and may therefore provide an example of dynamical
evolution during the pre-main-sequence phase.
\end{abstract}


\section{Introduction}

The optically undetectable source, WL~20 \citep[also known as BKLT
J162715\(-\)243843 and GY~240, see][for other aliases]{bklt}, was discovered
in a near-infrared (near-IR) bolometer survey of a \( 10\arcmin \times
10\arcmin \) region of self-absorbed \( ^{13} \)CO emission in the \( \rho \)
Ophiuchi star-forming cloud \citep{wl}. Soon thereafter, WL~20 was detected at
10~\microns\ from ground-based observations \citep{lw}, and at longer
wavelengths in the pointed observations mode of the \textit{IRAS} satellite,
where it is referred to as YLW~11 \citep{ylw}.

In the currently accepted classification scheme of young stellar objects
(YSOs) devised by \citet{lada}, WL~20 was one of the first sources to be
identified as a Class I source \citep{wly}. Empirically, Class I sources have
broader than blackbody spectral energy distributions (SEDs), with rising
2--10~\micron\ spectral slopes, \( a>0.3 \) (where \( a=\frac{d\log \lambda
F_{\lambda }}{d\log \lambda } \)). Theoretically, the SEDs of Class I sources
are interpreted to correspond to a remnant infalling dust and gas envelope
surrounding a central protostar\( + \)disk system \citep{als}. The more
evolved Class II sources, with \( -0.3>a>-1.6 \), are pre-main-sequence (PMS)
star\( + \)disk systems that have dispersed their remnant infall envelopes.

With the advent of near-IR array detectors, WL~20 was soon resolved into a
binary system at 2.2~\microns\ using a pixel scale of 0\farcs85 \citep{RAB}
and first reported to have an east-west separation of \( \approx \) 2\farcs7
from observations with a pixel scale of 0\farcs78 \citep{Bar89}. This
separation corresponds to \( \approx \) 340~AU for an adopted distance to the
cloud of 125~pc.

We note that some confusion still exists as to the distance to the \( \rho \)
Ophiuchi clouds, primarily due to many authors citing the distance to the
adjacent Sco-Cen OB association instead. The distance to the Sco-Cen OB
association has been determined to be 160\( \pm \)10~pc photometrically
\citep{whit} or 145~pc astrometrically \citep{deZ99}. By contrast, the
distance to the \( \rho \) Ophiuchi cloud itself has been determined to be
125\( \pm \)25~pc from detailed kinematic studies of the cloud gas
\citep{deG89,deG92}, and, more recently, by a study of \textit{Hipparcos}
parallaxes and Tycho \( B-V \) colors of stars of Classes III \& V, which show
an abrupt rise in reddening at \( d=120 \)~pc, as expected for a molecular
cloud \citep{knu}. We therefore adopt a distance of 125~pc to the cloud in
this paper.

The first indication that WL~20 is a triple system came from a deep ProtoCAM
survey to identify near-IR counterparts of centimeter continuum sources in the
\( \rho  \) Ophiuchi cloud core \citep{sks}. The VLA source identified with
WL~20 is known as LFAM~30, and was imaged at 6~cm with an 11\arcsec \( \times  \)5\arcsec\ beam
\citep{LFAM}. With the sensitive ProtoCAM images acquired at a pixel scale of
0\farcs20, a third and weakest 2.2~\micron\ component of the system was easily
identified, and designated as {}``30S{}'' (for the Southern component of
LFAM~30), whereas the components of the previously known near-IR binary are
referred to as {}``30E{}'' and {}``30W{}'' by these authors \citep{sks}. With
30E as the positional reference, 30W is quoted at a separation of 3\farcs3 at
P.A. 269\degr, and 30S at a separation of 3\farcs9 at P.A. 232\degr.

Near-IR spectra of the two brighter \( K \) components of the WL~20 triple
system, WL~20:E (\( K=10.13 \)) and WL~20:W (\( K=10.40 \)) were presented as
part of a spectroscopic survey of YSOs with \( K<10.5 \) in \( \rho \) Oph
\citep{gl}. The near-IR spectra of both WL~20:E \& W were found to be
consistent with those of other Class II sources, with a K--M spectral type
established for WL~20:E, and a more precise K7--M0 spectral type determination
for WL~20:W, due to the less severe continuum veiling in its spectrum
\citep{gm}. At their assumed distance of 160~pc, WL~20:E was determined to
have a bolometric luminosity,
\( L_{bol}=0.4L_{\sun } \), through \( A_{V}=15.4 \), whereas WL~20:W was
determined to have \( L_{bol}=1.7L_{\sun } \) through \( A_{V}=18.1 \).
Continuum veiling in the spectrum of WL~20:E precluded its placement on the
H-R diagram, whereas the lesser veiling in the spectrum of WL~20:W allowed a
mass estimate of 0.3~\( M_{\sun } \) from its location along a \( 3\times
10^{5} \)~yr pre-main-sequence isochrone \citep{gm}.

A more recent spectroscopic survey of near-IR sources in the \( \rho \)
Ophiuchi core including many sources as faint as \( K\sim 12 \), was obtained
at higher spectral resolution (\( R=1200 \)) than the previously published
surveys (with
\( R\leq 1000 \)), allowing for significantly improved classifications for
G through M spectral types \citep{lr}. These authors assigned a K6 spectral
type to WL~20:E (GY 240B), with a bolometric luminosity of 0.55~\( L_{\sun }
\), and an M0 spectral type to WL~20:W (GY 240A) with \( L_{bol}=1.4L_{\sun }
\) (they also assumed a distance of 160~pc). Both sources were found to have a
foreground extinction of \( A_{V}=16.3 \). WL~20:S, with \( K\sim 12.6 \), was
excluded from this survey as well, due to its relative dimness at near-IR
wavelengths.

Interestingly, the discrepancy between the Class I SED classification of WL~20
on the one hand, and its near-IR Class II spectroscopic classifications on the
other, has not been remarked upon previously. This inconsistency can only be
addressed by producing spatially resolved SEDs of the individual components of
this triple system. Until now, the highest spatial resolution photometry of
WL~20 longward of 4.8~\microns\ has been through a 6--8\arcsec\
aperture---confusing all three components \citep{lw}. In order to better
constrain the properties of WL~20:S, and of the triple system of which it is a
member, we have obtained new, unprecedentedly high (sub-arcsecond) spatial
resolution, ground-based mid-IR images of the WL~20 system, at six separate
wavelengths spanning the 8--25~\micron\ atmospheric window. Additionally, we
present spatially resolved near-IR imaging of this triple system. Finally, we
have performed careful astrometry, allowing us identify the source of the
centimeter continuum emission.

\section{Observations}

\label{sec:obs}

All mid-infrared images were obtained with MIRLIN, JPL's 128\( \times \)128
pixel Si:As camera. Diffraction-limited images were obtained on the nights of
1996 April 24 at the Palomar 5-m telescope, 1998 March 13--14 on the Keck II
10-m telescope, and 2000 June 16 at NASA's 3-m Infrared Telescope Facility
(IRTF). Pixel scales of MIRLIN were 0\farcs15 at Palomar, 0\farcs138 at Keck
II, and 0\farcs475 at the IRTF. Observations at Palomar and the IRTF were made
with the broadband N filter (\( \lambda =10.8 \)~\microns, \( \Delta \lambda
=5.7 \)~\microns). For reference, the full-width at half maximum (FWHM) of a
diffraction limited image at N-band is 0\farcs47 at Palomar and 0\farcs78 at
the IRTF. Observations at the Keck II telescope were made with narrower
filters, with central wavelengths (bandwidths) of 7.9~\microns\
(0.76~\microns), 10.3~\microns\ (1.01~\microns), 12.5~\microns\
(1.16~\microns), 17.9~\microns\ (2.00~\microns), 20.8~\microns\
(1.65~\microns), and 24.5~\microns\ (0.76~\microns); corresponding FWHMs range
from 0\farcs17 at 7.9~\microns\ to 0\farcs53 at 24.5~\microns.

The flux of WL~20 at each wavelength was determined by comparison with \(
\alpha \) Sco at Palomar, and at Keck II with a combination of \( \alpha \)
Boo, \( \alpha \) CMa, \( \alpha \) CrB, \( \alpha \) Hya, \( \beta \) Leo,
and \( \sigma \) Sco, the last of which proved to be an easily resolved
0\farcs45 binary. The weather at the IRTF was sufficiently poor that no flux
standards were observed. Data were obtained with traditional mid-IR chopping
and nodding techniques. The raw images were background-subtracted, shifted,
and coadded with our in-house IDL routine {}``MAC{}'' (Match-and-Combine).
Photometry for the standard stars was performed in 2\farcs5 and 2\farcs3
diameter apertures for the Palomar and Keck II data, respectively, when the
separation between components was adequate (as determined by the intensity
contours falling to zero between the sources) and by a combination of aperture
summation and PSF-fitting when the separation between WL~20:W and WL~20:S was
not clean. Typically, the flux ratio of a given component with respect to
WL~20:E was measured via PSF-fitting, then WL~20:E was calibrated with respect
to the standards. The photometric consistency between all the standards at
Keck II was found to be 5--10\% at all wavelengths. Though \( \alpha \) Sco
was the only standard observed at Palomar, multiple observations of it over
the course of the night are consistent to 5\% and observations of most other
WL sources \citep{wl} observed on that night agree with published values to
that level.

Although the IRTF data were taken under extremely non-photometric conditions,
special care was used during the course of the observations to acquire
accurate astrometry, in order to establish which of the three components of
the WL~20 system is responsible for the observed radio continuum emission
\citep[hereafter, LFAM]{LFAM}. We imaged three additional nearby Class I radio
emitters---LFAM~23 (WL~22), LFAM~27 (WL~15), and LFAM~33 (YLW~15) in turn with
LFAM~30 (WL~20), with as little delay between images as possible, so that
registration of the mid-IR source location with respect to the radio
coordinates could be made as accurately as possible. If each VLA source were
exactly positionally coincident with its respective mid-IR counterpart, then
each mid-IR source should fall on exactly the same pixel in each MIRLIN image.
In order to reduce our astrometric errors, seven images were obtained of
WL~15, five of WL~20, three of WL~22, and two of YLW~15. The resulting scatter
in the source positions for WL~22 and YLW~15 was only 0.5~arcsec, allowing
relative astrometric determinations to this accuracy.

Near-infrared images of WL~20 were acquired with ProtoCAM, a 58\( \times \)62
InSb array on the NASA IRTF atop Mauna Kea, on the night of 1990 August 03, at
1.29, 1.67, 2.23, 3.55, and 3.82~\microns. The pixel scale of these
observations was 0\farcs35. The data were not photometric: thin cirrus clouds
were present throughout the observations. Software aperture photometry
(4\farcs5 apertures) was performed on the standards (HD147889 and BD 65\( +
\)1637) and the fluxes were found to be consistent only at the 30--50\% level
over a fairly wide range of airmasses. Nevertheless, even under such
conditions, the \textit{intensity ratios} between the components of WL~20
(\textit{i.e.}, the flux ratio of WL~20:W/WL~20:E and of WL~20:S/WL~20:E)
should be unaffected by weather because the sources all fall within the
instantaneous field-of-view.

\section{Results}

\subsection{Imaging, Photometry, and SEDs}

\begin{figure}
[!t]

{\par\centering \resizebox*{0.95\columnwidth}{!}{\includegraphics{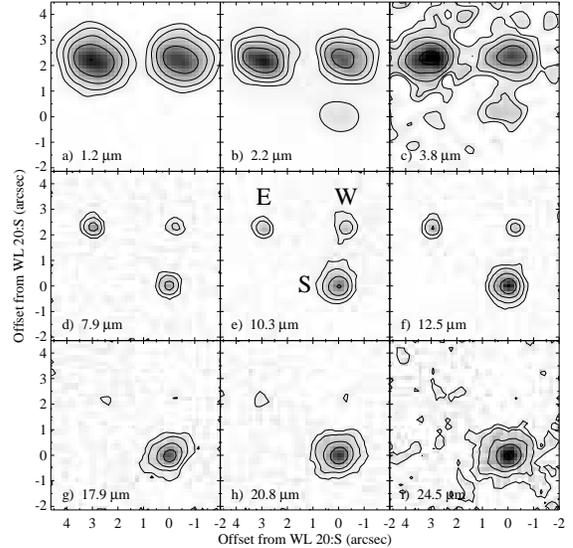}} \par}

\caption{Images of the WL~20 triple system at near- and mid-infrared wavelengths. (a--c)
Images of the WL~20 system acquired with ProtoCAM at the IRTF
(0\farcs35/pixel, resampled to 0\farcs138/pixel); (d--i) MIRLIN Keck II images
(0\farcs138 pixel scale). Contours in each panel represent the flux density in
Jy~arcsec\protect\( ^{-2}\protect \), and are spaced by one magnitude. The
lowest contour in panel \emph{a} is 400
\protect\( \mu \protect \)Jy; in \emph{b} it is 4 mJy; in \emph{c--f}, 40 mJy;
and in \emph{g--i}, 160mJy. The individual system components, WL~20:E, WL~20:W,
and WL~20:S are identified in \emph{e}. The field of view of each panel is
6\farcs6\protect\( \times \protect \)6\farcs6. East is to the left, and north
is up for all of the images presented in this paper. Source separations are
3\farcs17\protect\( \pm \protect \)0\farcs01 between WL~20:E and WL~20:W,
and 2\farcs26\protect\( \pm \protect \)0\farcs02 between WL~20:W and WL~20:S.\label{fig:img}}
\end{figure}

We present diffraction-limited (\( \sim \) 0\farcs25 resolution at
10~\microns) mid-infrared images of the WL~20 triple system acquired with
MIRLIN at the Keck II telescope, along with representative shorter wavelength
images acquired with ProtoCAM at the IRTF in Figure~\ref{fig:img}. We list the
source separations and position angles derived from mean positions obtained
from the 7.9, 10.3, and 12.5~\micron\ images in Table~\ref{tab:sep}. The
individual components of the WL~20 triple system are labeled in
Figure~\ref{fig:img}\emph{e} (the 10.3~\micron\ image). It is evident from
inspection of Figure~\ref{fig:img} that whereas WL~20:S is the weakest source
of the system at the shortest wavelengths, it gradually brightens towards the
longer wavelengths, just as its companions to the north are dimming.
Eventually, WL~20:S dominates the system luminosity at the longest wavelengths
(17.9, 20.8, and 24.5~\microns).

We have used a combination of software aperture summation and point-spread
function fitting to obtain photometric information for each component of the
WL~20 system individually. Fluxes derived from the new data presented here, as
well as all known previously published values, are listed in
Table~\ref{tab:fluxes}.

\begin{figure*}[!t]

\begin{center}

\resizebox*{0.95\columnwidth}{!}{\includegraphics{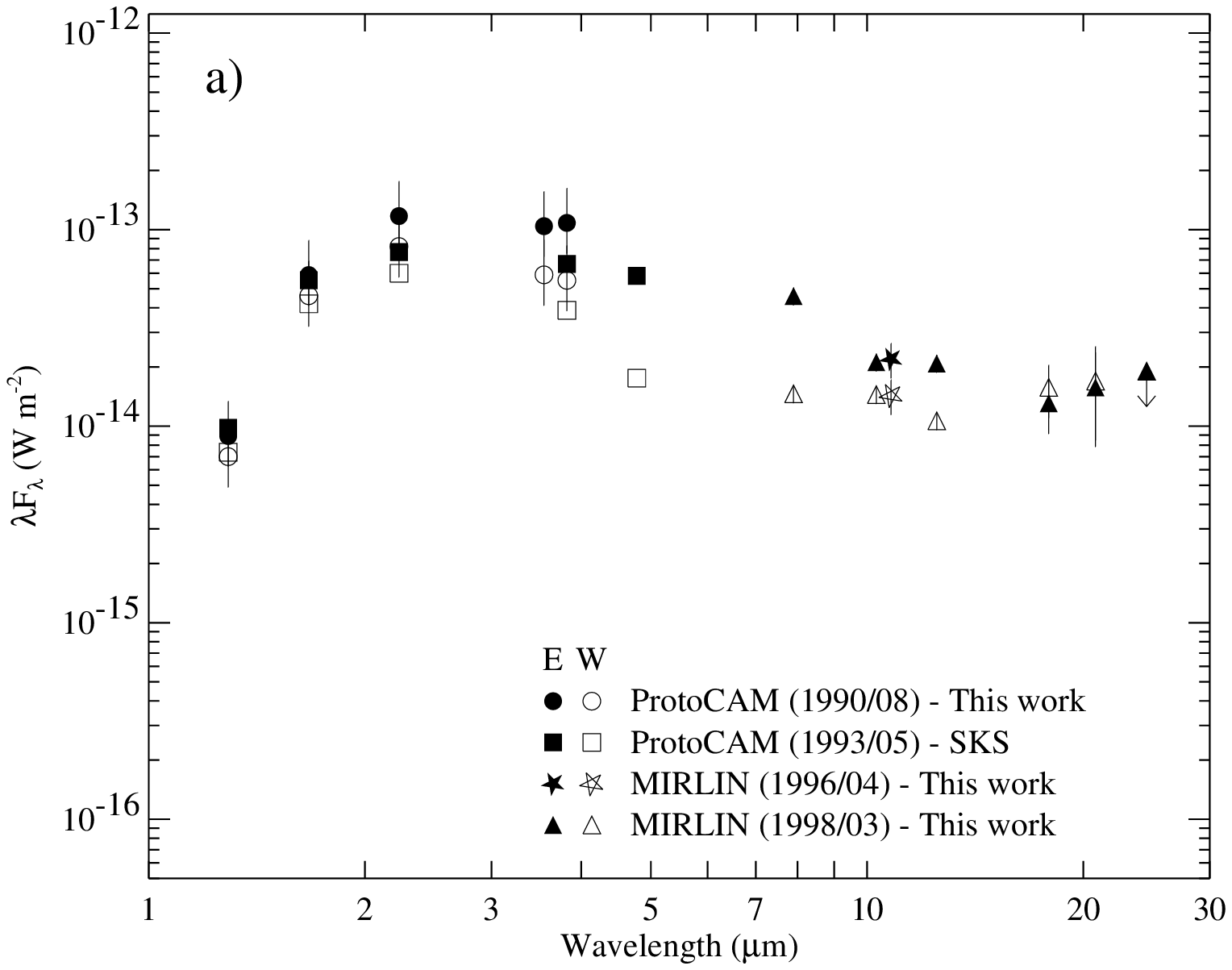}} \hspace*{0.4in}\resizebox*{0.95\columnwidth}{!}{\includegraphics{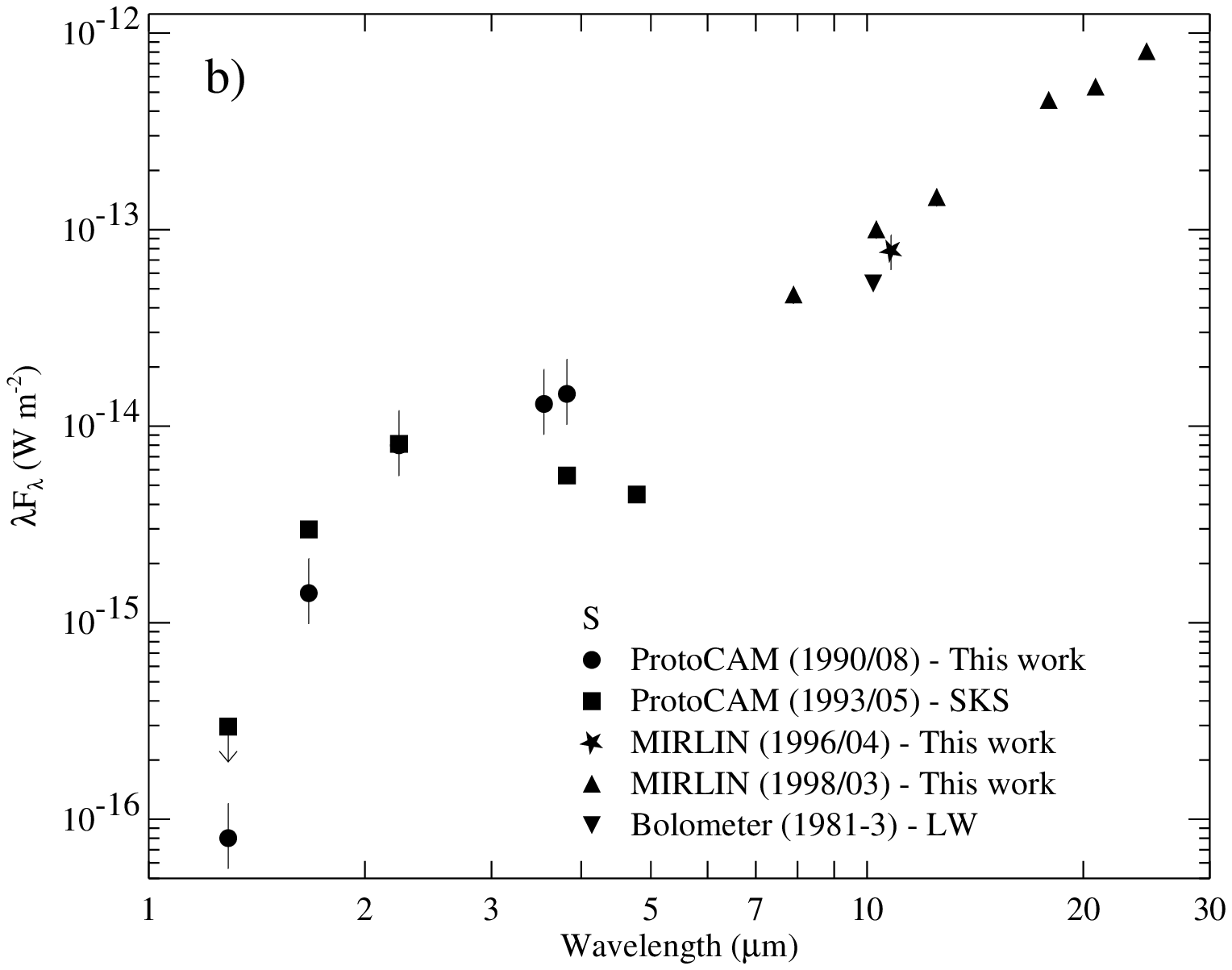}} 

\end{center}

\caption{Spatially-resolved SEDs of the three components of the WL~20 system.
(a) The spectral energy distributions of the east and west components of
WL~20. Filled symbols represent photometry for WL~20:E, open symbols show
photometry for WL~20:W. Squares are ProtoCAM data from \citet{sks}, circles
are ProtoCAM data from 1990 August, stars indicate our N-band MIRLIN
photometry from 1996 April Palomar 5-m observations, and triangles indicate
mid-infrared MIRLIN photometry from 1998 March Keck II observations. The
negative 2--10~\micron\ spectral slopes characteristic of Class II objects are
evident for each of these two sources. (b) The spectral energy distribution of
WL~20:S plotted to the same scale as \emph{a}. Symbols are as in \emph{a},
except that filled symbols now represent WL~20:S, and the total N-band system
flux from \citet{lw} has been added (inverted triangle). The SED of WL~20:S is
of a completely different character than either WL~20:E or WL~20:W, and is
similar to that of younger Class I sources.}\label{fig:sed_ir}

\end{figure*}

\begin{figure}
[!t]

{\par\centering \resizebox*{0.95\columnwidth}{!}{\includegraphics{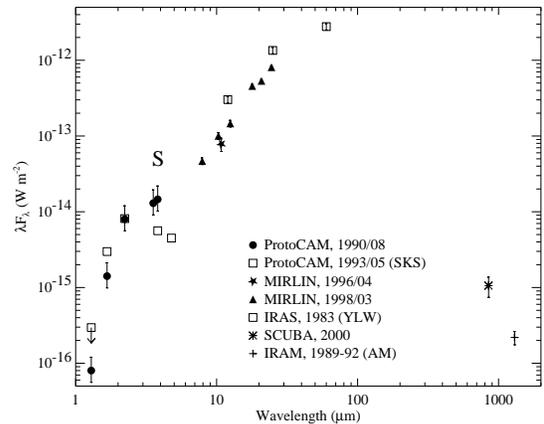}} \par}

\caption{The SED of WL~20:S from 1~\microns\ -- 1.3~mm. The full spectral energy
distribution of the southern component of WL~20 including \emph{IRAS}, SCUBA,
and millimeter data. (There is no \emph{IRAS} 100~\micron\ point due to the
severe confusion in this region.)\label{fig:sed_s}}
\end{figure}

The resulting spectral energy distributions (SEDs) are presented in
Figures~\ref{fig:sed_ir} and \ref{fig:sed_s}. The SEDs for E and W
(Figure~\ref{fig:sed_ir}\emph{a}) are consistent with their being reddened
Class II sources with modest excesses at long wavelengths, in agreement with
the near-IR spectroscopic results. The spectral slopes, \( a=-0.79 \) for
WL~20:E, and \( a=-0.91 \) for WL~20:W, are as expected for Class II sources.
In fact, the shape of the SED of WL~20:W is quite close to a reddened
blackbody in the near-IR (\emph{i.e.} a small near-IR excess, though the
excess at mid-IR wavelengths is substantially larger). Thus it is possible
that the circumstellar material around WL~20:W may be optically thin enough
for silicate emission to be present. Comparison of the 10.3~\micron\ fluxes
(near the center of the silicate feature) with the {}``continuum{}'' fluxes at
7.9 and 12.5~\microns\ suggests a factor of \( \sim 2 \) excess at
10.3~\microns\ in WL~20:W with respect to that anticipated from the shape of
WL~20:E's SED. Our single data point is not sufficiently compelling (though it
is robust) to warrant a large discussion of silicate emission here, but future
spatially-resolved mid-IR spectroscopy should address in fine detail the
nature of the dust emission and absorption in these Class II sources.

\begin{table*}[!t]

\caption{Relative position of the three components.\label{tab:sep}}
\smallskip{}

\begin{tabular}{cccc}
\hline 
Pair &
 Separation &
 Separation &
 Position Angle\\
&
 (arcsec) &
 (AU) &
 (\degr)\\
\hline 
W with respect to E&
 3.17\( \pm  \)0.01&
 400&
 270.1\( \pm  \)0.3\\
 S with respect to E&
 3.66\( \pm  \)0.03&
 460&
 232.2\( \pm  \)0.2\\
 S with respect to W&
 2.26\( \pm  \)0.02&
 280&
 173.0\( \pm  \)0.3 \\
\hline 
\end{tabular}

\end{table*}

\begin{table*}[!t]

\caption{Fluxes for the WL~20 components.\label{tab:fluxes}}
\smallskip{}

\renewcommand{\arraystretch}{0.8}

\begin{tabular}{cccccccccc}
\hline 
&
\multicolumn{3}{c}{This Work}&
\multicolumn{3}{c}{SKS\tablenotemark{a}}&
 This Work&
 SKS\tablenotemark{a}&
 Other\\
 Wavelength&
 E&
 W&
 S&
 E&
 W&
 S&
 Total&
 Total&
 Total\\
 (\microns)&
 (mJy)&
 (mJy)&
 (mJy)&
 (mJy)&
 (mJy)&
 (mJy)&
 (mJy)&
 (mJy)&
 (mJy)\\
\hline 
1.29&
 3.8\tablenotemark{*}&
 3.0\tablenotemark{*}&
 0.03\tablenotemark{*}&
 4.2&
 3.2&
 \( < \) 0.1&
 6.8\tablenotemark{*}&
 7.5&
 6.9\tablenotemark{b}\\
 1.67&
 32.7\tablenotemark{*}&
 25.6\tablenotemark{*}&
 0.79\tablenotemark{*}&
 30.7&
 23.3&
 1.7&
 59.1\tablenotemark{*}&
 55.8&
 50.6\tablenotemark{b}\\
 2.23&
 87.1\tablenotemark{*}&
 61.0\tablenotemark{*}&
 5.9\tablenotemark{*}&
 57.1&
 44.6&
 6.0&
 154\tablenotemark{*}&
 108&
 125\tablenotemark{b}\\
 3.55&
 123.2\tablenotemark{*}&
 69.7\tablenotemark{*}&
 15.3\tablenotemark{*}&
&
&
&
 208\tablenotemark{*}&
&
 158\tablenotemark{b}\\
 3.82&
 137.7\tablenotemark{*}&
 70.3\tablenotemark{*}&
 18.6\tablenotemark{*}&
 85.1&
 49.4&
 7.1&
 227\tablenotemark{*}&
 142&
\\
 4.78&
&
&
&
 92.6&
 28.0&
 7.2&
&
 128&
\\
 6.7&
&
&
&
&
&
&
&
&
160,240,290\tablenotemark{c}\\
 7.9&
 121.&
 38.4&
 123.&
&
&
&
 282&
&
\\
 10.2&
&
&
&
&
&
&
&
&
180\tablenotemark{d}\\
 10.3&
 72.6&
 49.6&
 345.&
&
&
&
 467&
&
\\
 10.8&
 79.0&
 51.5&
 281.&
&
&
&
 412&
&
\\
 11.5&
&
&
&
&
&
&
&
&
630\tablenotemark{c}\\
 12&
&
&
&
&
&
&
&
&
1200\tablenotemark{e}\\
 12.5&
 86.8&
 44.3&
 610.&
&
&
&
 741&
&
\\
 17.9&
 78.0&
 93.9&
 2720.&
&
&
&
 2900&
&
\\
 20.8&
 109.&
 117.&
 3700.&
&
&
&
 3930&
&
\\
 24.5&
 \( < \)155.&
 \( < \)155.&
 6600.&
&
&
&
 6910&
&
\\
 25&
&
&
&
&
&
&
&
&
11200\tablenotemark{e}\\
 60&
&
&
&
&
&
&
&
&
55600\tablenotemark{e}\\
 850&
&
&
&
&
&
&
&
&
300\tablenotemark{f}\\
 1300&
&
&
&
&
&
&
&
&
95\tablenotemark{g} \\
\hline 
\end{tabular}

\tablenotetext{*}{Non-photometric.}

\tablerefs{a) \citet{sks}; b) \citet{wl}; c) \citet{wilk}; d) \citet{lw}; e)
\citet{ylw}---data have been color-corrected as described in the text; f)
\citet{scuba}; g) \citet{am}}

\end{table*}

The SED of WL~20:S (Figure~\ref{fig:sed_ir}\emph{b}), however, is that of a
Class I source, with \( a=+1.44 \). Given that there is a generally continuous
slope between our ground-based mid-IR data for WL~20:S and the far-infrared
and millimeter fluxes for the entire WL~20 system, we have attributed all the
longer wavelength flux observed in this system to WL~20:S
(Figure~\ref{fig:sed_s}). For the case of the millimeter emission, this
assumption may be tested by future interferometric observations.

We color-correct the \emph{IRAS} fluxes for this source in order to make the
best luminosity estimate possible. We use our narrowband observations with
MIRLIN at 12.5 and 24.5~\microns\ along with others' 850~\microns\ and 1.3~mm
observations to constrain the shape of the far-IR SED so the color-correction
terms may be estimated. The spectral slope implied by all the narrowband
mid-IR data follows a \( F_{\nu }\propto \nu ^{-3} \) power law over this
range, so we deredden the 12 and 25~\micron\ fluxes according to that rule
from the \emph{IRAS Explanatory Supplement} \citep{expsupp}, dividing them by
factors of 0.91 and 0.89, respectively. The SED clearly peaks in the vicinity
of 60~\microns\ (or at least does not rise significantly throughout the entire
45--80~\micron\ passband). We therefore fit the 60, 850, and 1300~\micron\
data with a \( \sim \)~80~K greybody. In fact, the 60~\micron\ color
correction at this temperature is quite small (divide by 0.97) and is not very
sensitive to modest excursions in temperature (60--120 K), so that even using
the uncorrected flux would be adequate. The resulting data show the usual
factor of 2--3 excess of the \emph{IRAS} measured flux vs. the ground-based
data \citep[\emph{e.g.}, ][]{lw}.

\subsection{Luminosities of the Individual Components of WL~20}

\label{sec:lum}

\begin{figure}
[!t]

{\par\centering \resizebox*{0.95\columnwidth}{!}{\includegraphics{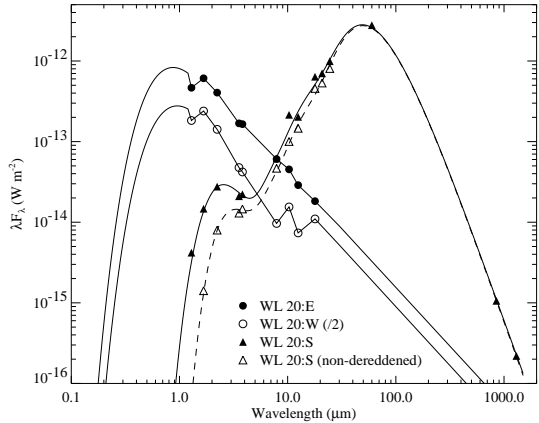}} \par}

\caption{Luminosity determinations for the three components of the WL~20 system. The
data for the three components, along with the curves used to obtain the best
luminosity estimates, are plotted. The data for WL~20:W have been divided by
two for clarity. The data for all three components have been dereddened by
assuming an \protect\( A_{V}=16.3\protect \) foreground screen and a
\citet{dl} extinction curve (solid lines). An non-dereddened curve for WL~20:S
(dashed line) is also plotted for the case where all the extinction is local
to the source and dereddening is not appropriate. The {}``model{}'' curves for
WL~20:E \& W are represented directly by the data from 1--18~\microns, a
blackbody of the appropriate effective temperature (see the text) from 0.1 to
1.2~\microns, and a power law beyond 18~\microns. The curves for WL~20:S are
composites of several blackbodies modified for extinction and dust emissivity.
The curves for WL~20:S assume all far-IR and millimeter flux originate from
WL~20:S; in the case that only 50\% does (see the text), the 60~\micron\ and
millimeter points are reduced by a factor of two and the model curves
\protect\( >\protect \) 40~\microns\ are correspondingly
reduced.\label{fig:lum}}
\end{figure}

We compute the luminosities of each Class II component of the WL~20 system (E
and W) by integrating under the curves displayed in Figure \ref{fig:lum}. The
data points from \( 1.2\leq \lambda \leq 18 \)~\microns\ have been dereddened
using a \citet{dl} extinction curve assuming \( A_{V}=16.3 \) \citep{lr}.
Though we assume \( A_{V}=16.3 \), we can rule out \( A_{V}>18 \) from our
data since the near-IR data then have a slope steeper than a blackbody, and \(
A_{V}<15 \) can be ruled out as the near-IR data would then be too red to be
consistent with the flux from a \( \sim \) 4000~K photosphere. We use a power
law extrapolation from 18 through 1300 microns consistent with the spectral
slope from 3--18~\microns\ (\( F_{\nu }\propto \nu ^{0.45} \)) to estimate the
source fluxes at far-infrared through millimeter wavelengths. Other
extrapolations for the fluxes from 18~\microns\ to longer wavelengths may be
used (\emph{e.g.} no flux at all past 18~\microns, or even a constant flux
between 18 and 1300~\microns), but none change the estimated luminosity more
than 2\%, since the majority of the energy is emitted at shorter wavelengths.

A strict lower limit to the source luminosities, the {}``infrared{}''
luminosities, are arrived at by integrating under the dereddened
1--18~\micron\ data points and the long wavelength extrapolation described
above using simple trapezoidal integration. We find an infrared luminosity of
0.26~\( L_{\sun } \) for WL~20:E, and 0.17~\( L_{\sun } \) for WL~20:W. We
have used our near-IR photometry to compute the luminosity rather than that of
\citet{sks} since this provides better continuity with the mid-IR data.
However, use of the \citet{sks} near-IR data reduces our luminosity estimates
by only 0.03~\( L_{\sun } \) for each component.

To obtain a more accurate value of the total luminosity of each component,
however, we scale a blackbody spectrum from 0.1--1.2~\microns\ at the
effective temperatures found by \citet{lr} to match the observed infrared data
values. \citet{lr} found temperatures of 4205 K for WL~20:E and 3850 K for
WL~20:W. We then integrate under this blackbody curve in addition to the
infrared data points (the solid curves in Figure \ref{fig:lum}). This
technique yields total luminosities of 0.61~\( L_{\sun } \) for WL~20:E and
0.39~\( L_{\sun } \) for WL~20:W. Integration of only the blackbody (over all
wavelengths), which should approximate the emission of the photosphere without
luminosity from the disk, yields 0.53 and 0.35~\( L_{\sun } \), respectively.
These values are appropriate for comparison with the pre-main sequence
evolutionary models discussed in Section \ref{sec:pmsmodels}.

For WL~20:S, computing the luminosity is complicated by two issues: if and how
to deredden the data, and how to partition the flux at far-IR and millimeter
wavelengths where the sources are not resolved. With regard to dereddening,
since no photospheric measurements are available, we can assume either that
all the extinction is local and most of the near-IR photons are absorbed and
reradiated in the far-IR where the extinction is much smaller, or that the
same \( A_{V}=16.3 \) foreground screen is present as for the other two
components (and there is a correspondingly lower local extinction).
Dereddening the data in the first case would {}``double count{}'' the near-IR
photons, while dereddening in the second is the correct procedure. However,
since most of the energy is radiated in the mid-IR through millimeter regime
where the extinction is small, the final luminosity estimate changes only 10\%
after dereddening.

As for the far-IR and millimeter fluxes, we can establish an upper bound on
the luminosity by assuming that all the 60~\micron\ \emph{IRAS} flux from
\citet{ylw}, the 850~\micron\ flux from SCUBA \citep{scuba}, and the 1.3~mm
measurement from \citet{am} originates from WL~20:S. As a lower bound, we
assume that 50\% of the flux originates from WL~20:S; this is consistent with
the ground-based 12 and 25~\micron\ points being roughly half the \emph{IRAS}
points. In either case, this necessarily implies that most of the dust in the
WL~20 system surrounds WL~20:S. Also, the \emph{IRAS} data presumably contain
some contaminating flux from the nearby source WL~19; however, we know that
WL~19 is quite faint with respect to WL~20 at 10.8~\microns\ \citep[factor of
5.6 fainter, ][]{brc}; therefore its effect on our luminosity estimate will be
very small.

With all the above factors in mind, a simple trapezoidal integration of the
WL~20:S fluxes yield a bolometric luminosity of 1.28~\( L_{\sun } \) if all
the extinction is local, and 1.40~\( L_{\sun } \), if the data points are
first dereddened for an \( A_{V}=16.3 \) foreground screen. In both these
instances, all the observed far-IR and millimeter fluxes were assigned to
WL~20:S. If, instead, we assign only half the observed far-IR and millimeter
flux to WL~20:S, the corresponding values for the luminosity become 0.84~\(
L_{\sun } \) and 0.95~\( L_{\sun } \), respectively. Use of the \citet{sks}
near-IR data, instead of our near-IR data, increases these values by only
0.04~\( L_{\sun } \).

To improve the above luminosity estimates for WL~20:S, given the coarseness of
the trapezoidal integration algorithm and the sparseness of the data between
25 and 850~\microns, where the SED peaks, we have constructed a smooth model
curve which passes through all of the data points under which to integrate.
This curve is constructed by assuming a \( T_{eff}=4000 \)~K photosphere, two
blackbodies at \( T=300 \)~K and \( T=150 \)~K to represent the disk emission,
and a modified \( T=65 \)~K blackbody dust envelope with a \( \lambda ^{-1} \)
emissivity law. This model was created only to obtain a smooth curve which
represents the SED of WL~20:S adequately for integration; it is not intended
to be a physical description of the object. The corresponding derived
luminosities for WL~20:S then become 1.71~\( L_{\sun } \) for the curve
passing through the data points, 1.82~\( L_{\sun } \) for the curve dereddened
by \( A_{V}=16.3 \), assuming all of the far-IR and millimeter flux to
originate from WL~20:S, and 1.04~\( L_{\sun } \) and 1.14~\( L_{\sun } \),
respectively, for the case when only half the observed far-IR and millimeter
fluxes are attributed to WL~20:S. Therefore, the true luminosity of WL~20:S
lies somewhere between 1.0--1.8~\( L_{\sun } \), making this source the most
luminous member of the system by a factor of \( \sim \) 2.

\subsection{Variability of the Class I Source, WL~20:S}

\begin{figure}
[!t]

{\par\centering \resizebox*{0.95\columnwidth}{!}{\includegraphics{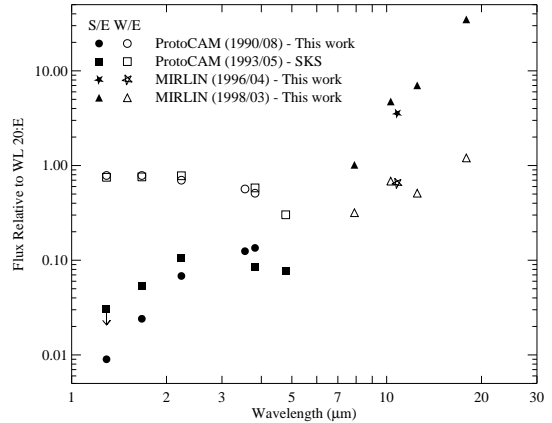}} \par}

\caption{SED variability of WL~20:S. Open symbols represent the ratio of the flux from
WL~20:W with respect to WL~20:E; filled symbols represent the ratio of the
flux from WL~20:S with respect to WL~20:E; otherwise, the symbols are the same
as in Figure~\ref{fig:sed_ir}\emph{a}. There was a significant shift in the
spectral shape of the Class I source, WL~20:S, between the 1990 and 1993
ProtoCAM observations (filled circles vs. filled squares), while the fluxes
and spectral shape of the Class II source, WL~20:W, over this wavelength range
remained essentially constant (open circles and squares). The change in the
WL~20:S/WL~20:E flux ratio at 10~\microns\ is also significant between the
1996 and 1998 MIRLIN observations.\label{fig:sed_rel}}
\end{figure}

The slight discrepancy of the measured flux in the broadband-N (10.8~\microns)
filter for WL~20:S in the April 1996 Palomar data from the flux measured for
this source with the the 10.3~\micron\ silicate filter in the March 1998 Keck
II data led us to examine the relative photometry of the components of the
WL~20 system as a function of time (see Table \ref{tab:fluxes}). For this
purpose, we replot the spatially resolved fluxes as intensity ratios with
respect to the fluxes of component WL~20:E in Figure~\ref{fig:sed_rel}. The
square symbols in this figure represent the previously published photometry
from \citet{sks} for data acquired in 1993 May. The circular symbols represent
the relative photometry from our 1990 August ProtoCAM observations. (Because
these are now relative ratios, the variations in the sky should divide out,
and we believe the relative photometric errors are less than 5\%.) The
different epochs of MIRLIN observations are represented by stars and
triangles. From this plot, it can be seen that whereas WL~20:W does not vary
significantly over this wavelength range and timescale, WL~20:S varies greatly
(factor of \( \sim 3 \)) in the near-IR and perhaps even changes the shape of
its SED over the timescale of a few years.

Even at 10~\microns, the variations appear to be significant. Though our
MIRLIN results show only a 25\% increase between 1996 and 1998, \citet{lw}
report a flux of 180 mJy for the \emph{entire system} (all three components
should have been contained in their 6~arcsec beam), whereas our total system
flux from 1998 is 470 mJy. If we assume WL~20:E \& W to be constant over this
entire timespan, WL~20:S would have had a flux of only 60 mJy in 1981--3,
resulting in a nearly 6-fold increase in 15 years.

\citet{wilk} have also reported that WL~20 brightened from 160 to 290 mJy at
6.7~\microns\ from 1996 through 1998. Piecing the observations together as
well as possible, it seems that WL~20:S was in a low luminosity state in the
early 1980s, brightened until \( \sim 1990 \), faded during the early 1990s,
then has been increasing since the mid-1990s. Given the changing shape of the
SED, it may be that as the accretion rate increases, both the luminosity and
the local extinction rise, leading to large increases in the mid- and
far-infrared and perhaps a decline in the near-infrared. Though much more
temporal data is required to put these speculations on a firm footing, they
seem plausible with the data in hand. We conclude that much of the total
luminosity of WL~20:S is derived from accretion, and that the stellar
luminosity cannot be directly determined from our data.

\subsection{Astrometry and the Identification of LFAM~30 with WL~20:S}

\begin{figure}
[!t]

{\par\centering \resizebox*{0.95\columnwidth}{!}{\includegraphics{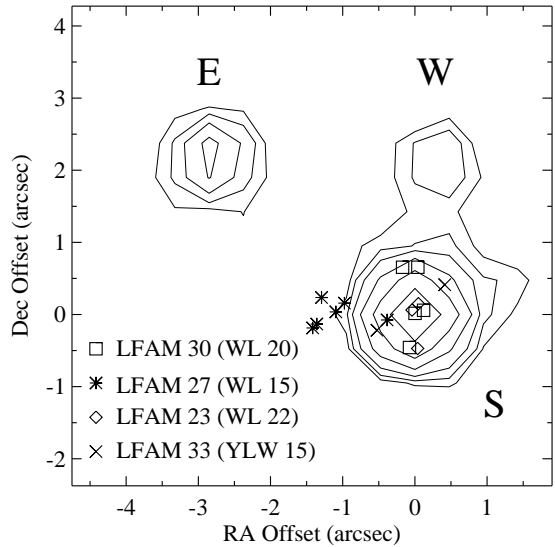}} \par}

\caption{Identification of LFAM~30 with WL~20:S. A plot of the offset positional errors
between LFAM~30 and WL~20 derived by successive offsetting between the cm and
mid-IR positions of four Class I radio emitters, indicated in the symbol key,
superposed on the 10.3~\micron\ contour plot of the WL~20 system. If we assume
that LFAM~23 and LFAM~33 are exactly coincident with their mid-infrared
counterparts, WL~22 and YLW~15, then LFAM~30 is coincident with WL~20:S. The
scatter in the relative positions is 0.5 arcsec. WL 15 appears to be offset
from the published radio position by about 1~arcsec in RA; but this is
consistent with the coarseness of the published VLA position (0.1 seconds of
time in RA). Even with this uncertainty, WL~20:E \& W are ruled out as
possible counterparts to the VLA source.\label{fig:offsets}}
\end{figure}

In their VLA survey of the the \( \rho \) Ophiuchi cloud, \citet{LFAM} found
that a number of cluster members were radio emitters. One of their radio
sources, LFAM~30, was associated with WL~20. However, until now, it had not
been possible to identify which of the three components of the WL~20 system is
responsible for the radio emission. In order to solve this problem, we
obtained the astrometric data at the IRTF as described in Section
\ref{sec:obs}. Assuming the mean position of LFAM~23 (WL~22) to be the zero
point, we plot the offsets from the \citet{LFAM} radio positions of these
sources superimposed upon the N filter image of WL~20 in
Figure~\ref{fig:offsets}. We therefore find that to within \( \pm
\)0.5~arcsec, LFAM~30 is coincident with WL~20:S.

WL~15 appears to be offset 1~arcsec east from the radio position. This is
almost certainly due to the fact that the VLA coordinates were published only
to the nearest 0.1 seconds of time, which is 1.4~arcsec on the sky at the
declination of WL~20, so an offset of 1~arcsec is not unexpected. However,
even if the total error were due to purely random pointing errors (which is
unlikely given the small scatter for the other objects), the consistency of
the radio offsets is sufficiently good that a correspondence between LFAM~30
and either E or W is ruled out.

\subsection{Extended Mid-Infrared Structure of WL~20:S}

\begin{figure}
[!t]

{\par\centering \resizebox*{0.95\columnwidth}{!}{\includegraphics{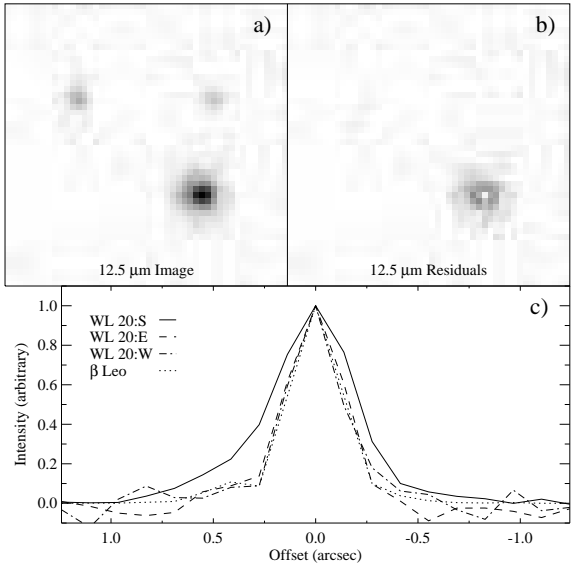}} \par}

\caption{The 12.5~\micron\ extent of WL~20:S. Figure \ref{fig:resid}\emph{a} shows
the gray-scale image of the WL~20 system at 12.5~\microns\ at a plate scale of
0.138~arcsec/pixel (6.6\protect\( \times \protect \)6.6~arcsec field-of-view).
Figure \ref{fig:resid}\emph{b} shows the leftover emission after subtraction
of appropriately scaled (to the peak flux value) images of \protect\( \beta
\protect \) Leo, a point source, placed at the positions of WL~20:E, WL~20:W,
and WL~20:S. Scaled PSFs can completely account for all of the 12.5~\micron\
emission from WL~20:E \& W, as is demonstrated by the lack of residual flux
emission seen at their positions in Figure \ref{fig:resid}\emph{b}. This same
figure demonstrates clear evidence for extended 12.5~\micron\ flux associated
with WL~20 S after point-source subtraction. Figure \ref{fig:resid}\emph{c}
shows E--W cuts across the point-source calibrator, \protect\( \beta \protect
\) Leo, WL~20:E, WL~20:W, and WL~20:S. Only WL~20:S has a broadened core and
wings, indicative of emission more extended than a point
source.\label{fig:resid}}
\end{figure}

\begin{figure}
[!tb]

{\par\centering \resizebox*{0.95\columnwidth}{!}{\includegraphics{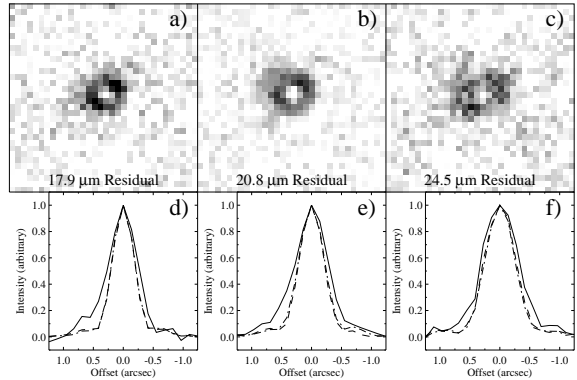}} \par}

\caption{The 17.9, 20.8, and 24.5~\micron\ extent of WL~20:S. Panels \emph{a--c} show
the result of subtracting a scaled PSF from the images at 17.9, 20.8, and
24.5~\microns, respectively. Panels \emph{a--c} cover a 4\farcs4\protect\(
\times \protect \)4\farcs4 field-of-view. Excess, extended emission is plainly
visible at each wavelength and is perhaps elongated in a
southeast-to-northwest orientation. Panels \emph{d--f} plot E--W intensity
crosscuts through WL~20:S (solid line) and two PSF calibrators (\protect\(
\alpha \protect \) CMa, dashed line, and \protect\( \alpha \protect \) Hya,
dot-dashed line). In all cases, the extended source size of WL~20:S appears to
be a constant \protect\( 45\pm 3\protect \)~AU, rather than a varying source
size which increases with increasing wavelength.\label{fig:cuts}}
\end{figure}

During the course of obtaining PSF-fit fluxes for each component of the WL~20
system, we discovered that WL~20:S is extended at all mid-infrared
wavelengths. In order to demonstrate this, we present images of the source
after the subtraction of a scaled (to the peak intensity value) PSF, as well
as intensity cross cuts along an E--W axis at 12.5, 17.9, 20.8, and
24.5~\microns\ in Figures~\ref{fig:resid} and \ref{fig:cuts}. Also shown for
comparison are similarly obtained intensity cross-cuts of the flux standards
used for the observations at each wavelength. Whereas the flux standards are
unresolved point sources, as are the two Class II sources, WL~20:E \& W,
WL~20:S is seen to have a definite extent above that expected for a point
source at each of the four plotted wavelengths. Somewhat surprisingly, the
physical size of the emitting region is fairly constant at all four
wavelengths (Table~\ref{tab:size}). The root-mean-square diameter (observed
FWHM -- PSF FWHM) is about \( 0.36\pm 0.03 \)~arcsec or \( 45\pm 3 \)~AU at
the distance of WL~20. The source size does not increase significantly with
increasing wavelength as is common in many Class I YSOs.

\begin{table*}[!t]

\caption{Source size of WL~20:S\label{tab:size}}
\smallskip{}

\begin{tabular}{ccccc}
\hline 
Wavelength&
 WL~20:S FWHM&
 PSF FWHM&
 RMS Difference&
 Diameter\\
 (\microns)&
 (arcsec)&
 (arcsec)&
 (arcsec)&
 (AU)\\
\hline 
12.5&
 0.45&
 0.28&
 0.36&
 45\\
 17.9&
 0.53&
 0.38&
 0.37&
 46\\
 20.8&
 0.58&
 0.43&
 0.38&
 48\\
 24.5&
 0.60&
 0.51&
 0.32&
 40 \\
\hline 
\end{tabular}

\end{table*}

\section{Discussion}

\subsection{WL~20: An {}``Infrared Companion{}'' System}

Infrared companion systems are young binary or multiple systems in which one
of the members is significantly {}``redder{}'' than the other members of the
system. {}``Red{}'' in this context means that the companion is often very
faint or invisible at optical and perhaps even the shorter near-IR
wavelengths, but very often dominates the luminosity of the system in the
near- and mid-IR. A prototypical example of such a system is T~Tau, with a
projected separation of 0\farcs73, corresponding to \( \sim 100 \)~AU at the
source. The northern component, T~Tau~N, exhibits a typical Class II spectrum.
Although the SED of its companion, T~Tau~S, is much redder than that of a
typical T~Tauri star, suggesting localized, high extinction towards this
source, the SED nevertheless peaks in the near-infrared (\emph{i.e.}, at \(
\sim 5 \)~\microns), unlike a Class I SED, which peaks towards the
far-infrared. T~Tau~S is detected only at wavelengths \( \geq 2.2 \)~\microns,
although its bolometric luminosity exceeds that of T~Tau~N by a factor of 2.
Furthermore, T~Tau~S has been shown to be variable (at the level of 2 mag flux
increases in the IR) over a 5~yr. time interval \citep{ghez,gorham}. Large
near-IR variability is a characteristic trait of infrared companion systems
\citep{mathieu}.

From our spatially resolved SEDs of the individual components of the WL~20
triple system, we can confidently assert WL~20 to be a newly identified member
of the class of infrared companion systems. Table~\ref{tab:properties} lists
the properties of each individual source derived from this work, with the
spectral types and effective temperatures for WL~20:E \& WL~20:W from
\citet{lr}.

\begin{table*}

\caption{WL~20 Source Properties\label{tab:properties}}

\begin{tabular}{cccccccc}
\hline 
Source&
 RA (2000.0)\tablenotemark{a}&
Dec (2000.0)&
Sp. Index&
 IR Lum&
 Total Lum&
 \( T_{eff} \)&
 Type \\
 WL~20&
 16\( ^{\textrm{h}} \) 27\( ^{\textrm{m}} \)&
\( - \)24\degr 38\( ^{'} \)&
\( a \)&
\( L_{\sun } \)&
 \( L_{\sun } \)&
 K&
\\
\hline 
E&
 15\fs82&
 43\farcs4&
 \( - \)0.79&
 0.26&
 0.61&
 4205&
 K6 Class II\\
 W&
 15.63&
 43.4&
 \( - \)0.91&
 0.17&
 0.39&
 3850&
 M0 Class II\\
 S&
 15.65&
 45.6&
 \( + \)1.44&
 0.84--1.40&
 1.04--1.82&
 N.A.&
 Class I \\
\hline 
\end{tabular}

\tablenotetext{a}{Assumes the RA of LFAM~23 is precisely 16\(^{\textrm{h}}\) 23\(^{\textrm{m}}\) 57\fs50 (1950.0).}

\end{table*}

The possibilities usually cited to explain such {}``infrared companion{}''
systems include the following:

\setlength{\leftmargini}{1.3em}

\begin{enumerate}

\setlength{\itemindent}{0.0em}

\item \emph{A chance superposition of sources} It may be that WL~20:S is not
physically associated with WL~20:E \& W---it is a chance superposition and
therefore WL~20:S can be in any evolutionary state relative to WL~20:E \& W.
We must stress than we cannot rule out this possibility based on currently
available data. It is nevertheless well established that all three sources of
the WL~20 system are YSOs, and therefore, all are associated with the \( \rho
\) Ophiuchi cloud. The space density of embedded objects in this cloud
\citep[\(<\) 200/sq.~degree, ][]{KLB} is low enough that finding three YSOs
apparently separated by such small distances is very unlikely unless they are
physically associated. Milli-arcsecond astrometry in the near-IR over a
sufficiently long time interval could prove association definitively, either
by showing the sources to have a common space motion or a definite orbital
motion.

\item \emph{A non-coplanar system where the IR companion is viewed {}``edge-on{}''}
It has recently been shown that a Class II source whose flared disk, with its
surface heated by the stellar radiation field, could mimic the SED of a Class
I source if the disk is in a nearly edge-on orientation to our line-of-sight
\citep{cg}. Indeed, we argue below that WL~20:S probably does have a flared
disk, though the inclination is \( \sim \) 20\( ^{\circ } \) from edge-on.
However, WL~20:S exhibits two phenomena common to IR companion systems that
orientation effects cannot explain if it is simply another Class II object: at
1.0--1.8~\( L_{\sun } \) it is the dominant luminosity source in the system by
a factor of two, and it is highly variable at both near- and mid-IR
wavelengths. Neither of these would be expected if it were simply an
extinguished sibling of the other two sources. We therefore consider this
scenario unlikely.

\item \emph{A younger age for the IR companion} It may be that WL~20:S formed
significantly later than its companions, WL~20:E \& WL~20:W. The projected
separations between the components of the WL~20 system range from 280--460~AU
(see Table~\ref{tab:sep}). These separations correspond to sound crossing
times of order \( \sim \) 1--\( 2\times 10^{4} \)~yr. Given that typical
free-fall times of pre-collapse cores are of order \( 10^{5} \)~yr, the WL~20
system must have collapsed from a single cloud core. Based on this simple
dynamical argument, it is highly unlikely that the individual components of
the WL~20 system formed at different times.

\item \emph{The objects are coeval; they are in a physical configuration and
a certain phase of evolution in which most of the dust accretion is occurring
on one of the objects, rather than all three.} This is the most plausible possibility
for explaining the properties of the infrared companion, WL~20:S, in the WL~20
system. Indications are that the observed 1.3~mm flux from the WL~20 system
is centered on the cm source \citep{am}, which we have shown is associated with
WL~20:S (see Figure~\ref{fig:offsets}). The observed 1.3~mm flux of 95 mJy
\citep{am} translates to a circumstellar mass of 0.03 (0.06)~\( M_{\sun } \),
assuming optically thin emission from dust at temperatures \( T_{d}=30 \)
(50)~K and \( \kappa _{1.3\, \mathrm{mm}}=0.01 \) cm\( ^{2} \) gm\( ^{-1} \).
Infrared variability, such as observed in WL~20:S (see
Figure~\ref{fig:sed_rel}), signals the presence of active accretion
\citep[\emph{e.g.}, ][]{beck}. The combined picture of this system is
reminiscent of recent binary formation models
\citep[\emph{e.g.}, ][]{bb} in which one component can be bright and actively
accreting relative to a less luminous secondary component, depending on the
initial distribution of specific angular momentum in the collapsing
protostellar envelope relative to the orbital angular momentum of the system.
We consider such a scenario to be the most likely one to explain the
properties of the WL~20 triple system.

\end{enumerate}

\subsection{Testing Pre-Main-Sequence Tracks with the WL~20 System}

\label{sec:pmsmodels}

To date, only a handful of pre-main-sequence binary or multiple systems have
both spatially resolved spectroscopy and spatially resolved photometry over
a wavelength range as broad as presented here (1--25~\microns). With these
data, we have been able to independently infer the luminosities of WL~20:E
\& W to significantly higher accuracy (\( \sim  \) a few percent apart from
systematic effects discussed below) than has been possible previously (see
Table~\ref{tab:properties}). This coeval system, with well-determined
photospheric luminosities (Section
\ref{sec:lum}) and effective temperatures, provides a stringent test to distinguish
between currently available PMS evolutionary models. Figures
\ref{fig:tracks}\emph{a--d} show isochrones (solid lines) and isomass (dashed
lines) evolutionary tracks at the same scale for four different sets of PMS
models, \citet{baraffe}, \citet{dantona},
\citet{ps}, and \citet{siess}, respectively. In each figure, the solid square
denotes WL~20:E and the solid diamond denotes WL~20:W. Figure
\ref{fig:tracks}\emph{e} shows the range of parameter space over which PMS
tracks are customarily plotted
\citep[\emph{e.g.}, ][]{ps}. The solid square outline in this figure indicates
the restricted range of parameter space plotted in Figures
\ref{fig:tracks}\emph{a--d}, in order to emphasize the improved precision with
which the different sets of tracks can be compared using the WL~20 data.

\begin{figure*}[!t]

\resizebox*{0.9\columnwidth}{!}{\includegraphics{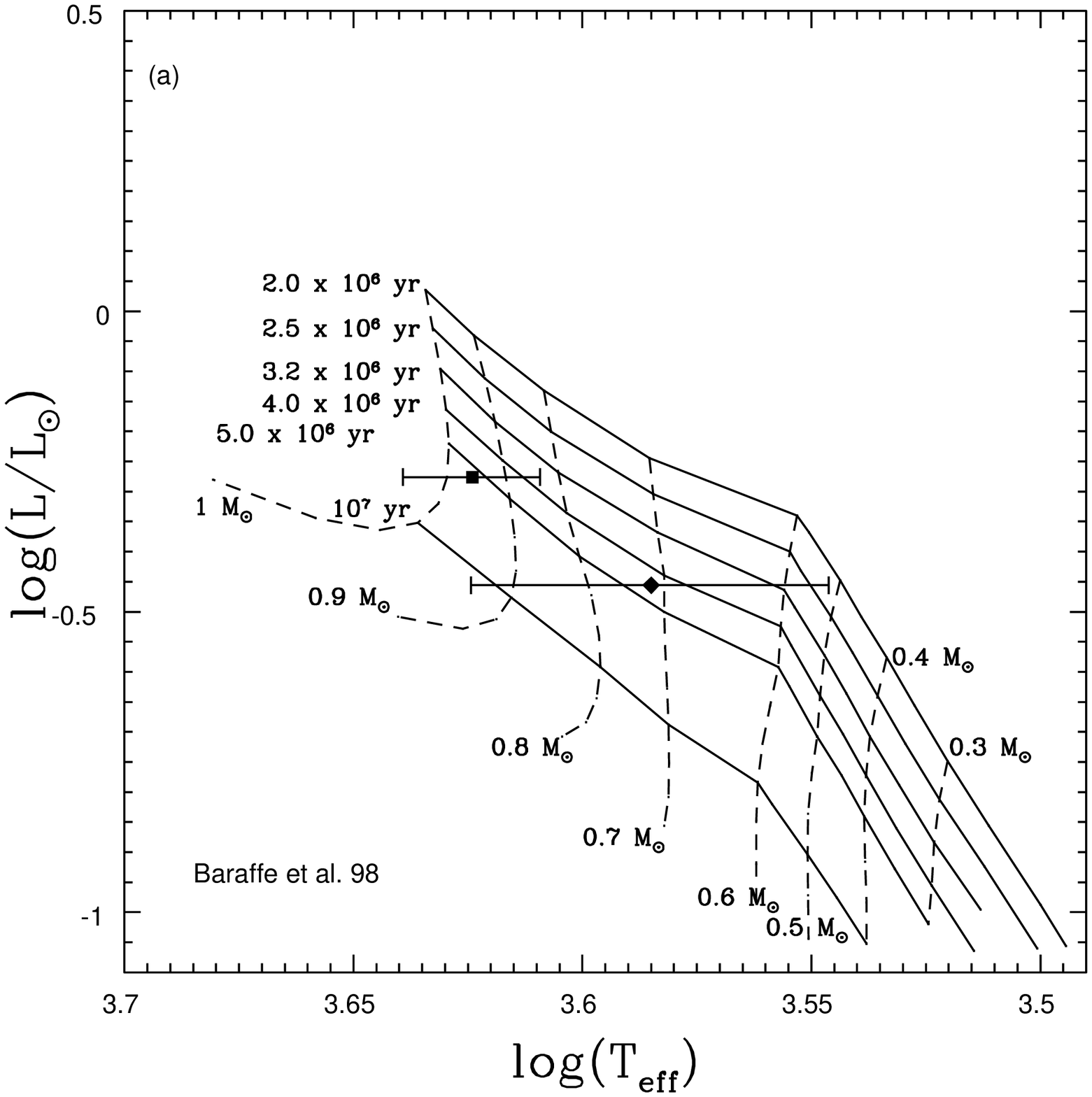}} \resizebox*{0.9\columnwidth}{!}{\includegraphics{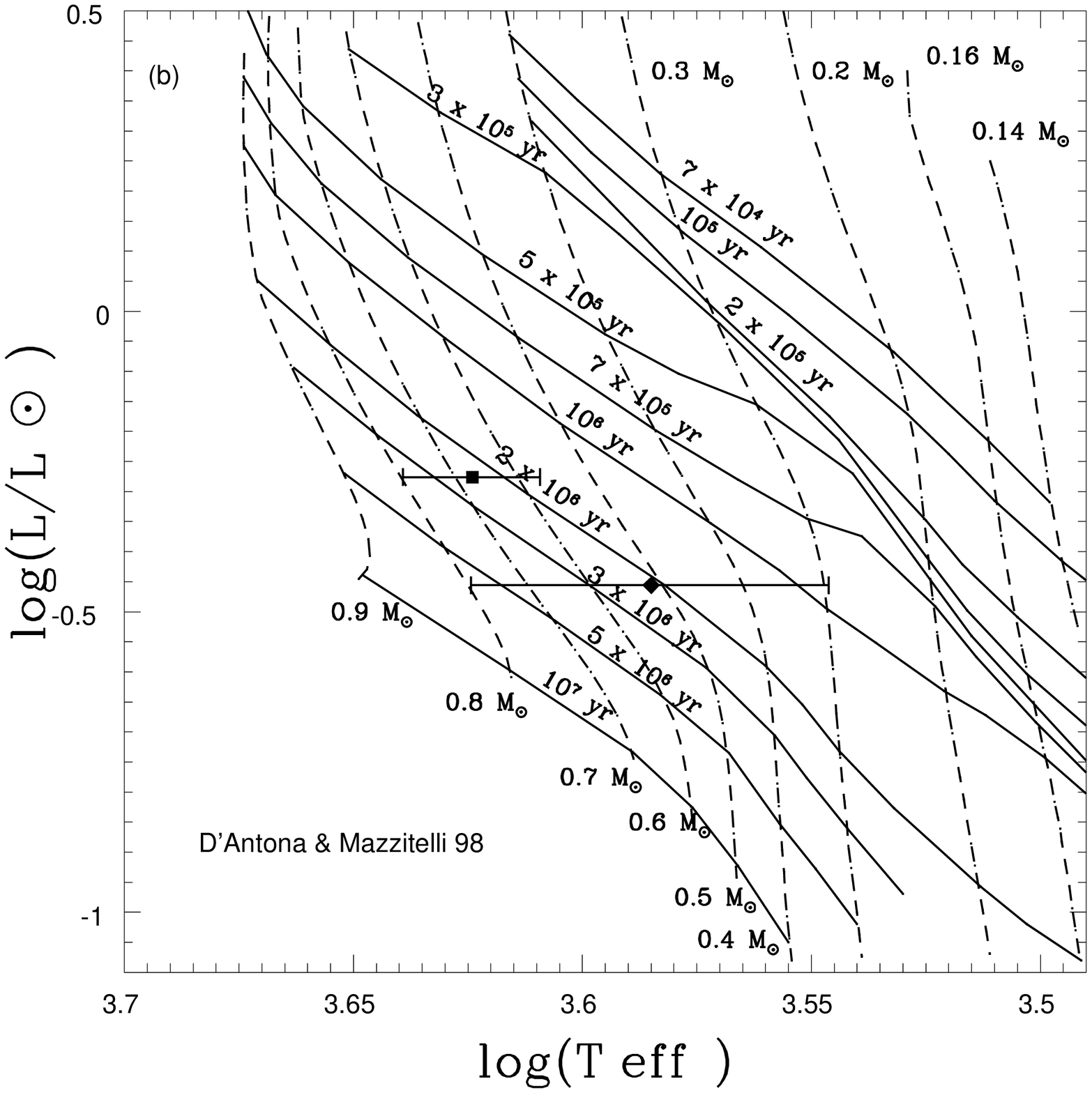}} 

\resizebox*{0.9\columnwidth}{!}{\includegraphics{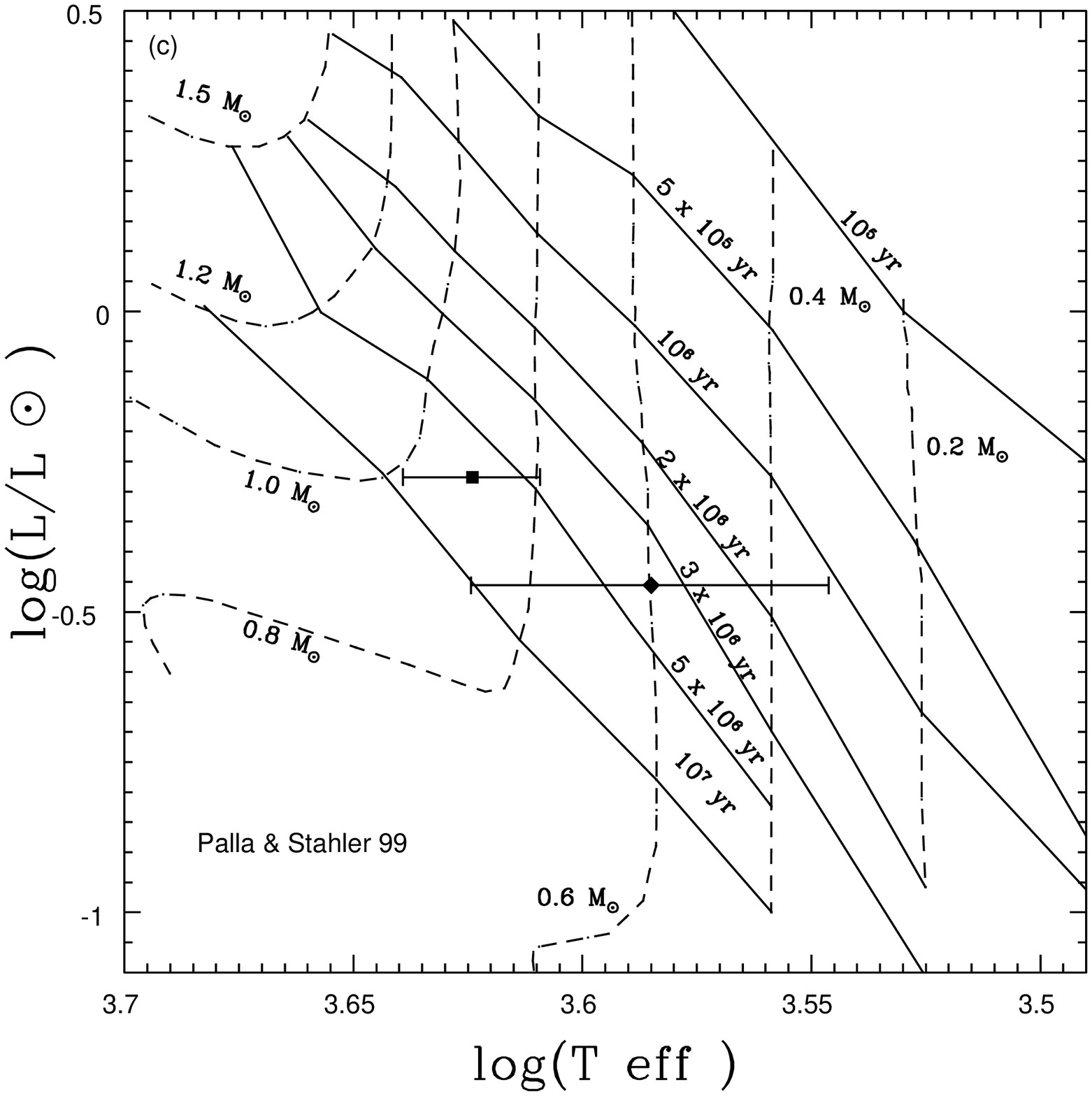}} \resizebox*{0.9\columnwidth}{!}{\includegraphics{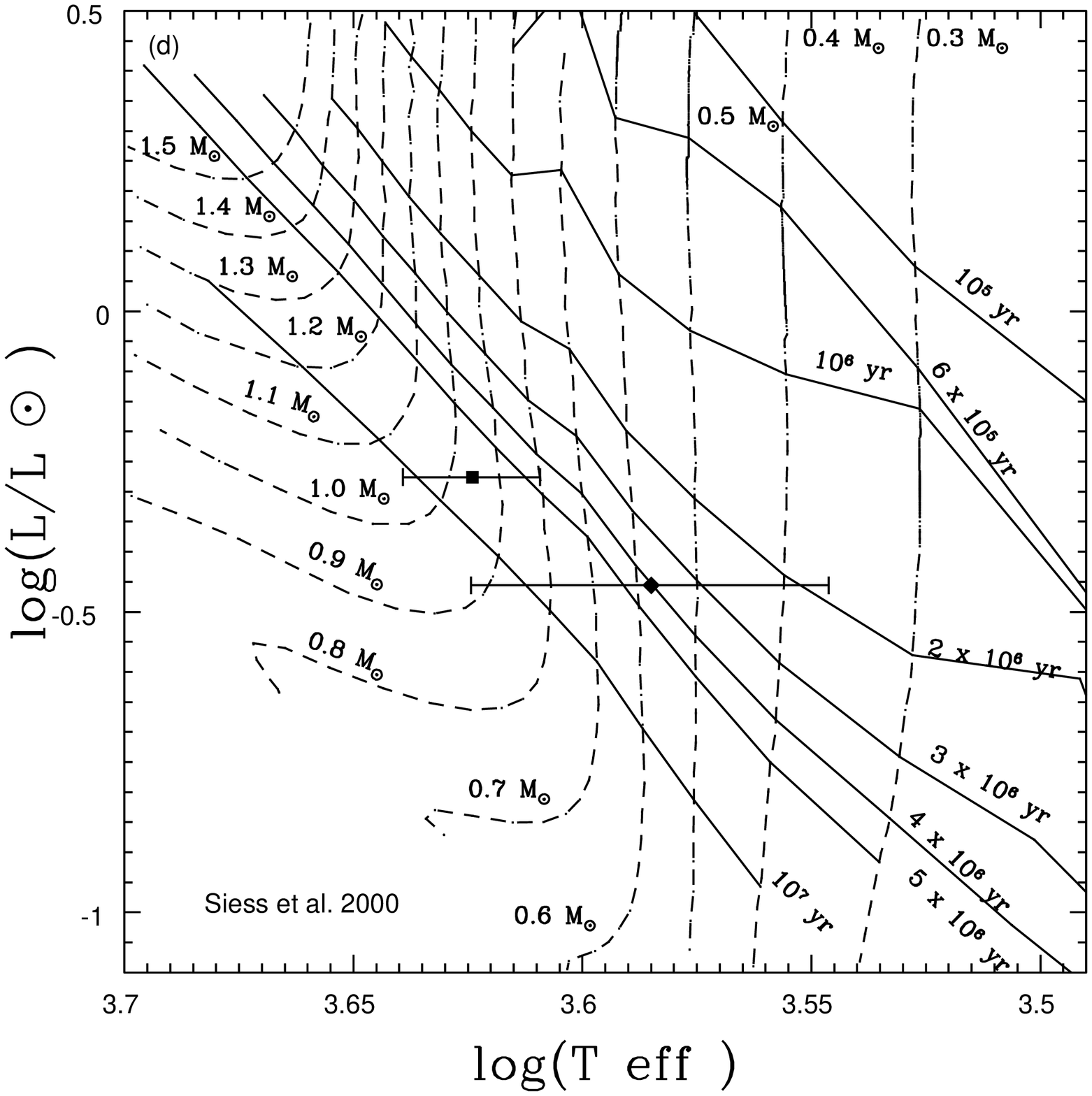}} 

\resizebox*{0.9\columnwidth}{!}{\includegraphics{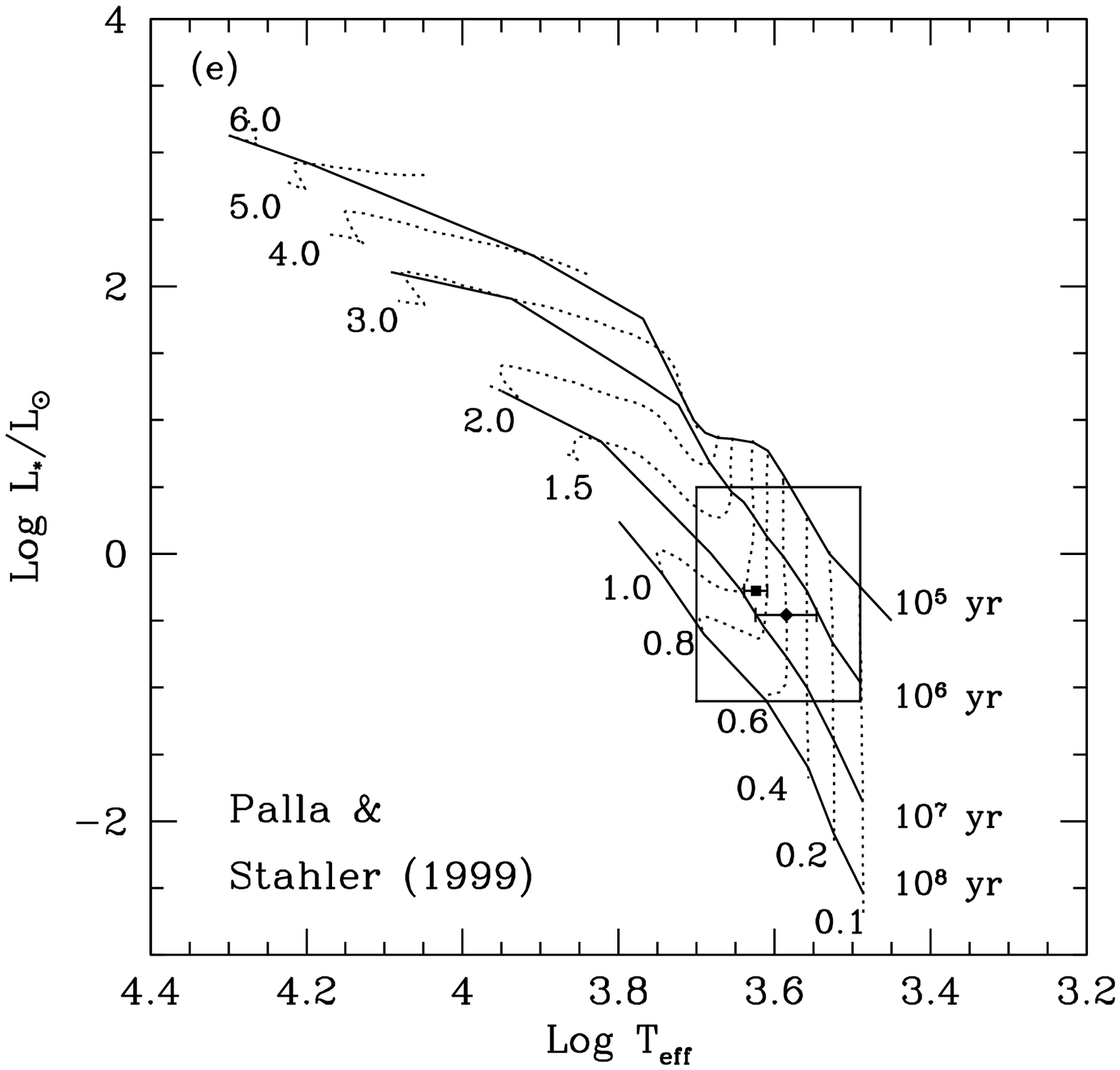}} 

\caption{\emph{See next page for caption.}}

\end{figure*}

\setcounter{figure}{8}

\begin{figure*}[!t]

\caption{Testing pre-main-sequence evolutionary tracks with the WL~20 system.
The first four panels of this figure display four separate sets of
pre-main-sequence evolutionary tracks over magnified luminosity and
temperature ranges. The stringency with which we are testing and comparing
these models is illustrated by the outlined box in Figure
9\emph{e}, which indicates the parameter ranges of the plots
presented in Figures 9\emph{a--d} compared with the scale of
previously plotted pre-main-sequence tracks \citep[\emph{e.g.}, ][]{ps}. In
all the panels, solid lines indicate isochrones, dashed lines indicate isomass
tracks, the filled square represents WL~20:E, and the filled triangle
represents WL~20:W. Figure 9\emph{a} shows the \citet{baraffe}
tracks. The youngest coeval system age still within the errors is $4\times
10^{6}$~yr. Figure 9\emph{b} shows the \citet{dantona}
tracks. The adopted temperatures and luminosities of WL~20:E \& W are
consistent with an age of 2.0--$2.5\times 10^{6}$~yr. This is the most
plausible (youngest) system age given by any of the tracks presented here, as
discussed in the text. Figure 9\emph{c} shows the \citet{ps}
tracks. The youngest system age, still within the data errors, and subject to
the coevality constraint is $\sim 5\times 10^{6}$~yr. Figure
9\emph{d} shows the \citet{siess} tracks. The youngest system
age, still within the data errors, and subject to the coevality constraint is
$\sim 5\times 10^{6}$~yr. Figure 9\emph{e} shows the usual
scale to which pre-main-sequence tracks are plotted. The rectangular area
outlines the plot limits of Figures 9\emph{a--d}, to highlight
the refined time resolution at which these tracks are tested by the WL~20
system.}\label{fig:tracks}

\end{figure*}

The two sources of systematic error in our luminosity determinations are the
adopted distance, for which we are using the Hipparcos-determined value, and
the adopted \( A_{V}=16.3 \). An error in the distance will move both sources
up by the same amount in each panel of Figure~\ref{fig:tracks}: a distance of
140~pc will raise the points by \( \sim 0.06 \) units in the log, or about one
small tick mark along the luminosity axis. Similarly, a slightly greater value
of extinction than the value of \( A_{V}=16.3 \) adopted here will also move
the position of each source up vertically in Figure~\ref{fig:tracks}. With an
extreme value of \( A_{V}=18 \), the luminosities are increased by
\( \sim 33 \)\%; 0.12 in the log, or 2.5 small tick mark(s).

From spatially resolved near-IR spectroscopy, \citet{lr} determine the
spectrum of WL~20:W to be consistent with a photosphere of spectral type
M2--K6, corresponding to a temperature range 3513~K \( \leq T_{eff}\leq \)
4205~K, with an adopted spectral type of M0 (corresponding to \( T_{eff}=3850
\)~K). The same authors assign a K6 (\( T_{eff}=4205 \)~K) spectral type to
WL~20:E, with possible spectral types in the range K5--K7, corresponding to
4060~K \( \leq T_{eff}\leq \) 4350~K. In Figure~\ref{fig:tracks}, we indicate
the possible effective temperature range for each source by the horizontal
error bars.

Inspection of Figure~\ref{fig:tracks} shows that none of the models rule out
WL~20:E \& W being a coeval pair, within the allowable errors in spectral type
for each source. However, the derived ages and masses differ amongst the
models. The \citet{dantona} tracks yield a system age of \( \sim \)~2.0--\(
2.5\times 10^{6} \)~yr, with a mass of 0.62--0.68~\( M_{\sun } \) for WL~20:E
and a mass of 0.51--0.55~\( M_{\sun } \) for WL~20:W. All the other models
yield a system age twice as old: 4--\( 5\times 10^{6} \)~yr. For an age of \(
4\times 10^{6} \)~yr, the \citet{baraffe} models yield masses of 0.86~\(
M_{\sun } \) for WL~20:E and 0.68~\( M_{\sun } \) for WL~20:W, respectively.
The \citet{ps} tracks yield a coeval system age of \( 5\times 10^{6} \)~yr,
with masses of 0.83~\( M_{\sun } \) for WL~20:E and 0.70~\( M_{\sun } \) for
WL~20:W. Similarly, the \citet{siess} tracks yield a coeval system age of \(
5\times 10^{6} \)~yr, with masses of 0.85~\( M_{\sun } \) for WL~20:E and \(
\sim 0.65M_{\sun } \) for WL~20:W.

Of the two possible ages for this system, 4--\( 5\times 10^{6} \)~yr or
2.0--\( 2.5\times 10^{6} \)~yr, the younger age is the more plausible one,
especially when we consider that WL~20:S, which appears to be a Class I
object, is also part of this system. Based on statistical arguments, Class I
objects are generally thought to be just a \( \mathrm{few}\times 10^{5} \)~yr
old \citep{wly}, maybe \( 8\times 10^{5} \)~yr at most \citep{khss}. Previous
statistical studies have ignored the systematic effects introduced by
unresolved binary/multiple systems, however, which have the effect of making a
source appear brighter (and therefore, judged to be younger), than is, in
fact, the case. Such an effect can result in derived ages of a factor of two
too young \citep{white}, so that the oldest Class I sources may be of order \(
1.6\times 10^{6} \)~yr old, just about consistent with the \( 1.8\times 10^{6}
\)~yr old system age derived from the \citet{dantona} and \citet{tout} tracks.

Very recently, pre-main-sequence tracks which include the effects of accretion
on the models have been calculated \citep{tout}. These authors provide the
magnitude of the errors possible when one derives ages and masses of PMS
objects from evolutionary tracks that ignore accretion, such as the ones
discussed above. In particular, placement of WL~20:E \& W on their Figure 14,
shows that the derived age from the other tracks, including those of
\citet{dantona}, can be up to a factor of two too old, relative to accreting
PMS models, making the true ages of WL~20:E and WL~20:W as young as 1--\(
1.3\times 10^{6} \)~yr, consistent with the oldest plausible Class I source
age of \( 1.6\times 10^{6} \)~yr for WL~20:S. The errors in the masses of PMS
objects derived from non-accreting vs.\ accreting tracks are much smaller,
being negligible in the case of WL~20:E and at the 10\% level for WL~20:W
\citep[found by placing these sources on Figure 13 of][]{tout}.

\subsection{The Nature of WL~20:S}

In this work, we have found that: 1) WL~20:S is the reddest and most luminous
member of the WL~20 system; 2) it is highly variable on timescales of a few
years; 3) it contains most of the dust in the system in a mid-IR emitting
region some 40~AU in diameter; and 4) that it is the source of the observed
centimeter emission.

The most likely explanation for the mid-IR appearance of WL~20:S is that it is
experiencing a phase of enhanced (and varying) accretion activity, perhaps due
to interactions with its neighbors. This is especially likely in view of the
fact that the SED of WL~20:S is consistent with the presence of a flared disk
of \( \sim 250 \)~AU in radius \citep[Eqn.~1]{cg}, very similar to the
projected separation of 280~AU between WL~20:S and WL~20:W. We observe a
structure whose 40~AU diameter size appears to be wavelength-independent in
the mid-IR: this structure may be the flared disk surface. The IRAS fluxes for
the WL~20 system, as a whole, are systematically larger than the sum of the
fluxes of the individual components derived from ground-based observations
(see Table \ref{tab:fluxes}). This discrepancy in the measured fluxes in
different-sized beams implies the presence of material on scales larger than
those to which the ground-based observations are sensitive, but that still
fall within an IRAS beam. Thus, \( \geq \) 40\% of the observed IRAS fluxes
are emitted from regions 10\( ^{\prime \prime } \)--120\( ^{\prime \prime } \)
in size, corresponding to the size scales of infalling envelopes. It may very
well be that WL~20:S is actively accreting matter from the envelope, through
its flaring disk, while its companions have already ceased significant
accretion.

If this is true, it addresses one of the primary objections to WL~20 being
a true triple system, as opposed to a chance superposition of a binary (WL~20:E
\& W) with a single source (WL~20:S). If WL~20 is a triple with an age of
\( \sim 2\times 10^{6} \)~yr, that is still uncomfortably old to have the
presence of a Class I source, which would normally be presumed to be \(
<1\times 10^{6} \)~yr old. This would appear to argue that WL~20:S is more
likely a chance superposition. However, if accretion has been continued to a
late phase due to tidal interactions with the other members, then the shape of
the SED of an individual source within a binary/multiple system is an
indicator only of the accretion activity of that source, and has little to do
with its age.

Preliminary studies of accretion in triple systems have so far focussed on
hierarchical triples, in which the separation between two sources is much
smaller than their distance to the third component, a circumstance clearly not
applicable to the WL~20 system. In fact, if formation proceeds through
fragmentation, then the resultant triples are typically not in a very
hierarchical configuration. The stability of accreting triples has been
examined by \citet{smith}. In general, if the maximum separation of the closer
pair (280~AU projected separation for WL~20:S \& WL~20:W) is comparable to the
minimum separation of this pair to the third component (400~AU projected
separation from WL~20:W to WL~20:E), then the stability of the system is
questionable. From the above analysis of existing PMS models, the current best
mass determinations for this system are 0.62--0.68~\( M_{\sun } \) for WL~20:E
and 0.51--0.55~\( M_{\sun } \) for WL~20:W, respectively, and \( \sim \)
1.0~\( M_{\sun } \) for WL~20:S from the constraints given by its 1.0--1.8~\(
L_{\sun } \) luminosity. According to the criterion for stability of triple
systems given by \citet{harr}, as quoted in \citet{smith}, the WL~20 system
should be dynamically unstable. The behavior of dynamically unstable accreting
triple systems has not yet been examined. It may be that some fraction of
single stars are formed from the disintegration of unstable triple systems.

\section{Conclusions}

We have presented sub-arcsecond, mid-infrared imaging photometry of the WL~20
triple system at 7.9, 10.3, 12.5, 17.9, 20.8, and 24.5 \( \mu \)m. When
supplemented by spatially-resolved, near-infrared imaging photometry from
ProtoCAM at the IRTF, these combined data allow solid determinations of the
spectral energy distribution of each source individually, as well as accurate
luminosity determinations. We find the source luminosities for WL~20:E,
WL~20:W, and WL~20:S to be 0.61~\( L_{\sun } \), 0.39~\( L_{\sun } \), and
1.0--1.8~\( L_{\sun } \), respectively. For WL~20:E and WL~20:W, 0.53~\(
L_{\sun } \) and 0.35~\( L_{\sun } \) can be attributed to photospheric
emission alone.

WL~20 can now be classified as an {}``infrared companion system,{}'' with
WL~20:S exhibiting an embedded protostellar (Class I) SED, while its two
neighbors, WL~20:E and WL~20:W, each exhibit T~Tauri star (Class II) SEDs. The
infrared companion, WL~20:S, is the dominant luminosity source in the system.
WL~20:S differs from its T~Tauri companions in three important respects: 1)
its near- and mid-IR fluxes vary significantly over timescales of years; 2) it
is well-resolved at mid-IR wavelengths, with a constant,
wavelength-independent source diameter of 40~AU; and 3) it is found to be the
source of the radio cm emission in the system.

Since the effective temperatures of WL~20:E \& W are known from
spatially-resolved near-IR spectroscopy, we can place these sources on a
Hertzsprung-Russell diagram. We are thus able to test currently available
pre-main-sequence evolutionary tracks at unprecedentedly high temporal
resolution and find that of the non-accreting models, the \citet{dantona}
tracks yield the most plausible system age, at 2.0--\( 2.5\times 10^{6} \)~yr.
The inferred source masses from these tracks at these ages are 0.62--0.68~\(
M_{\sun } \) for WL~20:E and 0.51--0.55~\( M_{\sun } \) for WL~20:W,
respectively.

We cannot, at present, independently determine a mass or age for the infrared
companion, WL~20:S. However, the intriguing possibility now exists of
determining the spectral type, and, therefore, the effective temperature of
this embedded source with the new generation of high-resolution spectrographs
on 8--10~m ground-based telescopes. Once an effective temperature
determination has been made spectroscopically, and assuming system coevality,
one could locate WL~20:S on an isomass track, independently of its known
luminosity. Thus, the possibility exists, for the first time, to directly
derive the accretion luminosity of a Class I protostar.

Millimeter interferometry of this unique triple system would advance our
understanding of the gas dynamics involved, processes which cannot be explored
in any other way. Higher temporal resolution spatially-resolved imaging and
monitoring of WL~20 at infrared wavelengths, combined with detailed modeling
of its appearance will also lead to a more detailed understanding of the
actual accretion processes taking place in WL~20:S.

\acknowledgements{}

We wish to thank Dr.~Michael Werner for a critical reading of this manuscript,
and Drs.~Bruce Wilking and Tom Greene for useful discussions and for providing
the impetus to publish these results. The anonymous referee provided a number
of comments, particularly in regard to the luminosity calculations, for which
we are grateful. We also thank Dr.~Wilking, Dr.~Derek Ward-Thompson and
Mr.~Jason Kirk for providing us with data in advance of publication and
Dr.~Russel White for providing computer-readable versions of certain sets of
pre-main-sequence tracks. MR thanks Dr.~Fred Chaffee and the entire Keck
Observatory staff for their enthusiasm, patience, and assistance in making it
possible to use MIRLIN on the Keck II telescope. The staffs of the Palomar
Observatory and the NASA IRTF have also continued to provide outstanding
support for MIRLIN as a visitor instrument on their respective telescopes.

Portions of this work were carried out at the Jet Propulsion Laboratory,
California Institute of Technology, under contract with the National
Aeronautics and Space Administration. Development of MIRLIN was supported by
the JPL Director's Discretionary Fund and its continued operation is funded by
an SR+T award from NASA's Office of Space Science. MB gratefully acknowledges
support from the National Science Foundation through grants AST-9731797 and
AST-0096087 to Harvey Mudd College, which have made her contributions to this
work possible.

\end{document}